\documentclass[reprint,notitlepage]{revtex4-1}

\usepackage{siunitx}
\usepackage{graphicx}
\usepackage{epstopdf}

\begin{document}
\title{Measurement of the $^{20,22}$Ne $^3\mbox{P}_2\mbox{--}^3\mbox{D}_3$ transition isotope shift using a single, phase modulated laser beam}

\author{B Ohayon, G Gumpel and G Ron}
\address{Racah Institute of Physics, Hebrew University, Jerusalem 91904, Israel}

\begin{abstract}
We develop a simple technique to accurately measure frequency differences between far lying resonances in a spectroscopy signal using a single, unlocked laser. This technique was used to measure the isotope shift of the cooling transition of metastable neon for the result of \SI{1626.287(53)}MHz. The most accurate determination of this value to date.
\end{abstract}


\maketitle

\section{Introduction}

Precise measurements of atomic optical transitions usually requires overcoming the large - typically few GHz - Doppler broadening of the lines, caused by the thermal distribution of the atoms.
To this end there exist a multitude of experimental techniques relying on either cooling (and/or trapping) of the sample, or limiting the interaction with probing fields to a specific, narrow velocity group. The latter method is generally called Doppler-free spectroscopy (DFS) \cite{1976-Hanch-DFS, 2013-Review-Percision}, and results in narrow lines, typically few MHz for optical transitions, which can be probed with a narrowband laser beam.
Whereas atomic-beam or trap setups require an elaborate vacuum system and sensitive detection for small observed signals, DFS of a thermal sample can be done with a gas sample in a cell, and enjoys large signal to noise ratio.
Finally, the systematic uncertainties in a vapor cell configuration are inherently different from those of cold atoms \cite{2015-QI_ACStrak, 2010-lith-sas}.

An accurate determination of the width of and interval between atomic resonances, requires calibration of the laser wavelength within the scanning range \cite{1977-SA_review}. A common way to achieve this is by using a cavity with known free-spectral-range (FSR) \cite{1971_Hansch_Lockin_pump}, which adds frequency markers in the form of narrow resonances whenever the laser is scanned over it. This method is limited by the uncertainty and drifts in the FSR, mostly due to thermal changes in the cavity length, and by scan linearity. To account for nonlinearity in the scanning procedure, many close markers are desired \cite{1980-FP-SA-Chooper-Rydberg, 2003-Li-EOM}, which require long cavities, that are more susceptible to thermal drifts.  Moreover, since the functional form of the nonlilnearity is generally unknown, and may change over time, interpolation errors may occur, which can be difficult to evaluate precisely.

A more elaborate method of calibrating the wavelength is to phase-lock a scan laser  to a reference laser locked to a stable feature, and measure their frequency difference. This method is limited by their noise, and the stability of the reference laser frequency during a measurement sequence. Higher stability is obtained when locking both lasers to a frequency comb \cite{2005-Helium-cell}, at the expense of a more elaborate and involved system.

Here we present a simple, versatile measurement scheme for precise determination of frequency differences between far lying resonances with different sizes. Our method overcomes most calibration challenges and drift errors, while using a single, unlocked laser. We demonstrate its applicability by measuring the isotope shift (IS) of the 
$2p^53s\mbox{ }^3\mbox{P}_2\mbox{ }(\SI{134041.8400}{cm^{-1}})
\stackrel{\rm{\SI{640.4}{nm}}}{\longrightarrow}
2p^53p\mbox{ }^3\mbox{D}_3\mbox{ }(\SI{149657.0393}{cm^{-1}})$ transition between $^{20}$Ne and $^{22}$Ne. A closed and  isolated transition used for laser-cooling applications \cite{1987-Shimizu}.

\section{Atomic signal with phase modulation}
\begin{figure}[b]
\includegraphics[clip,trim={130 5 280 10 },width=\linewidth]{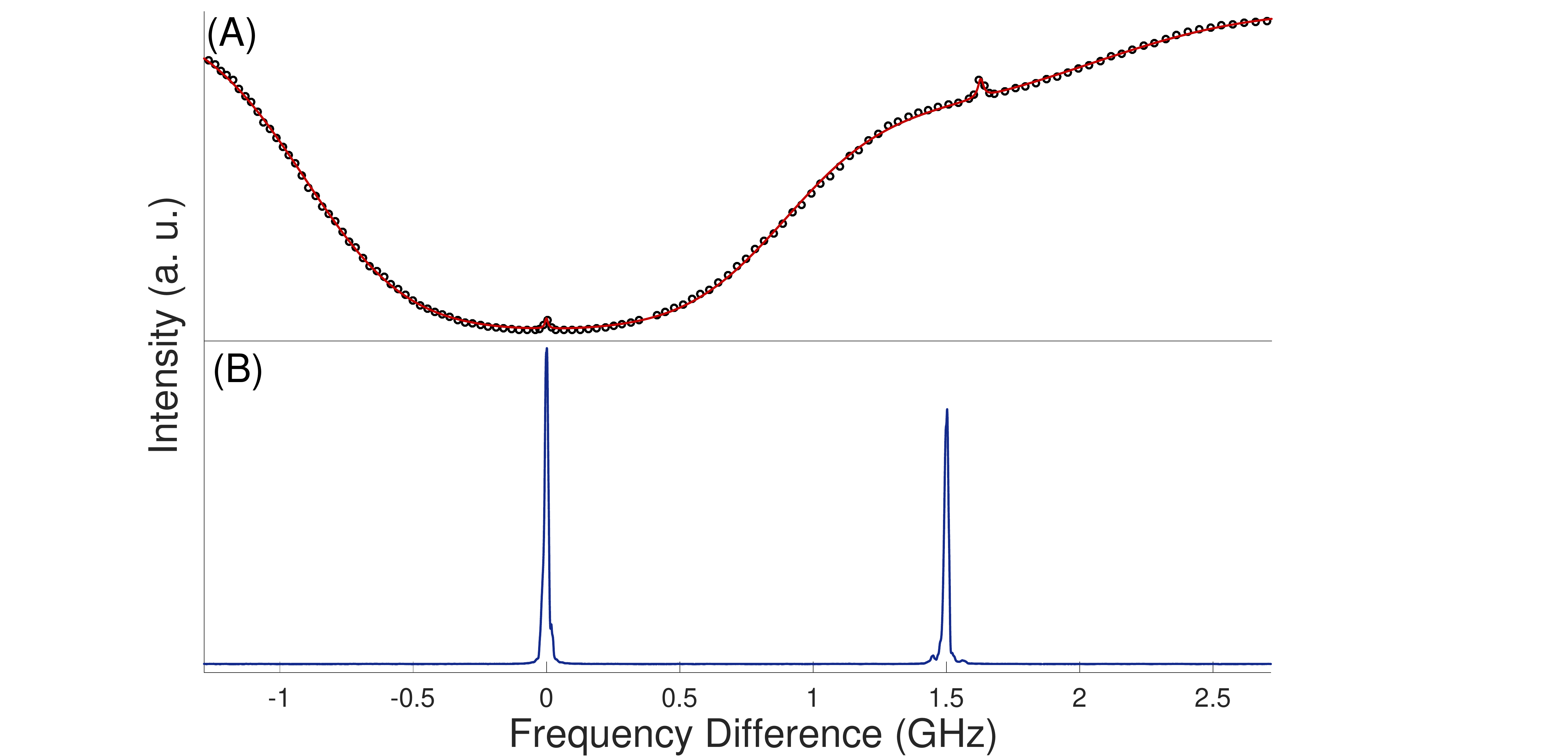}
	\caption{\label{Fig. Beer-Lambert}
		Wide scan of laser frequency without phase modulations or Doppler subtraction.
		(A) SA signal (circles), fitted with Eq. \ref{Trans_Two_exp} (solid line). Both $^{20}\mathrm{Ne}$ (left) and the less abundant $^{22}\mathrm{Ne}$ (right) Gaussian dips, and narrow Doppler-free peaks are observed.
		(B) FP signal.}
\end{figure}
We implement DFS by means of saturated absorption \cite{2013-Pressure}. For a single transition with a homogeneously broadened linewidth $\Gamma$, the transmission of a weak probe beam is approximated in the Doppler limit by \cite{1980_SA_Theory_Barium}
\begin{equation}
\mathrm{I}[\Delta]=I_0
 e^{
	-\mathrm{G}[\Delta](1-S \mathrm{L}[\Delta,\Gamma])
},
\end{equation}
with $\mathrm{G}$ the Doppler-broadened, Gaussian absorption coefficient of the atomic vapor, including the atomic density and cell length, $S$ is the resonance depth, which depends on the pump and probe intensities, and $\Delta$ the detuning from resonance. $\mathrm{L}$ is a normalized Lorentzian transmission function  $\mathrm{L}[\Delta,\Gamma]=1/(1+4(\Delta/\Gamma)^2)$.
For a sample containing two isotopes with an isotope shift of $\omega_{\mathrm{IS}}$, the transmission is given by
\begin{equation}\label{Trans_two}
\mathrm{I}[\Delta]=I_0
e^{
	-\mathrm{G}_1(1-S_1 \mathrm{L}_1) 
	-\mathrm{G}_2(1-S_2 \mathrm{L}_2)
}
\end{equation}
where, assuming that the transition in both isotopes has similar linewidth, 
$\mathrm{L}_2[\Delta] = \mathrm{L}_1[\Delta-\omega_{\mathrm{IS}}]              = \mathrm{L}[\Delta-\omega_{\mathrm{IS}},\Gamma]$,
$\mathrm{G}_2[\Delta] = \mathrm{G}_1[\Delta-\omega_{\mathrm{IS}}] n_2/n_1 = \mathrm{G}[\Delta-\omega_{\mathrm{IS}}] n_2/n_1,$
and we suppress notation of the frequency dependencies, $n_1,n_2$ are the isotopic atomic densities. We expand  (\ref{Trans_two}) in $\mathrm{G}_i S_i$  to get\begin{equation}\label{Trans_Two_exp}
\mathrm{I}[\Delta]=I_0
e^{-\mathrm{G}_1-\mathrm{G}_2}(1+\mathrm{G}_1S_1\mathrm{L}_1+\mathrm{G}_2S_2\mathrm{L}_2+O[(\mathrm{G}_i S_i)^2]).
\end{equation}
A trace of a broad frequency scan of the SA signal (without subtraction), fitted with   (\ref{Trans_Two_exp}), is shown in figure \ref{Fig. Beer-Lambert}.
When the pump beam is amplitude modulated with a frequency $\omega_\mathrm{c}<<\Gamma$ \cite{1971_Hansch_Lockin_pump,1980-FP-SA-Chooper-Rydberg}, the resonance depths are modulated as: $S_i\rightarrow S_i\cos(\omega_\mathrm{c} t+\phi)$. Feeding the modulated signal, along with the modulation, into a lock-in amplifier, the output in-phase component becomes:
\begin{equation}\label{two_lorenzians}
\mathrm{V}[\Delta] \propto \alpha_1\mathrm{L}_1+\alpha_2\mathrm{L}_2+O[(\mathrm{G}_i S_i)^3],
\end{equation}
where we evaluate the absorption coefficients on resonance: $\alpha_1=e^{-\mathrm{G}_1[0]-\mathrm{G}_2[-\omega_{\mathrm{IS}}]}\mathrm{G}_1[0]S_1,$ and $\alpha_2=e^{-\mathrm{G}_1[\omega_{\mathrm{IS}}]-\mathrm{G}_2[0]}\mathrm{G}_2[0]S_2$.
Equation (\ref{two_lorenzians}) describes two Loreznians on a flat background, separated by the isotope shift, with third order nonlinear corrections to the small peak amplitudes.
We now add phase modulation to the laser beam with a frequency much higher than the lock-in frequency $\omega_\mathrm{M}>>\omega_\mathrm{c}$. This creates sidebands in the pump and probe beams so that the resulting lock-in signal becomes:
\begin{equation}\label{crossovers}
\mathrm{V}[\Delta] \propto\\	
\sum_{a=-\infty}^{\infty}{\mathrm{J}_a^2
	(\alpha_1^a\mathrm{L}_1^a+\alpha_2^a\mathrm{L}_2^a) }+
\sum_{a\ne b}{\mathrm{J}_a\mathrm{J}_b
	(\alpha_1^{ab}\mathrm{L}_1^{ab}+\alpha_2^{ab}\mathrm{L}_2^{ab}) }, 
\end{equation}
\begin{figure}[b]
	\centering
	\includegraphics[clip,trim={220 12 140 10 },width=\linewidth]{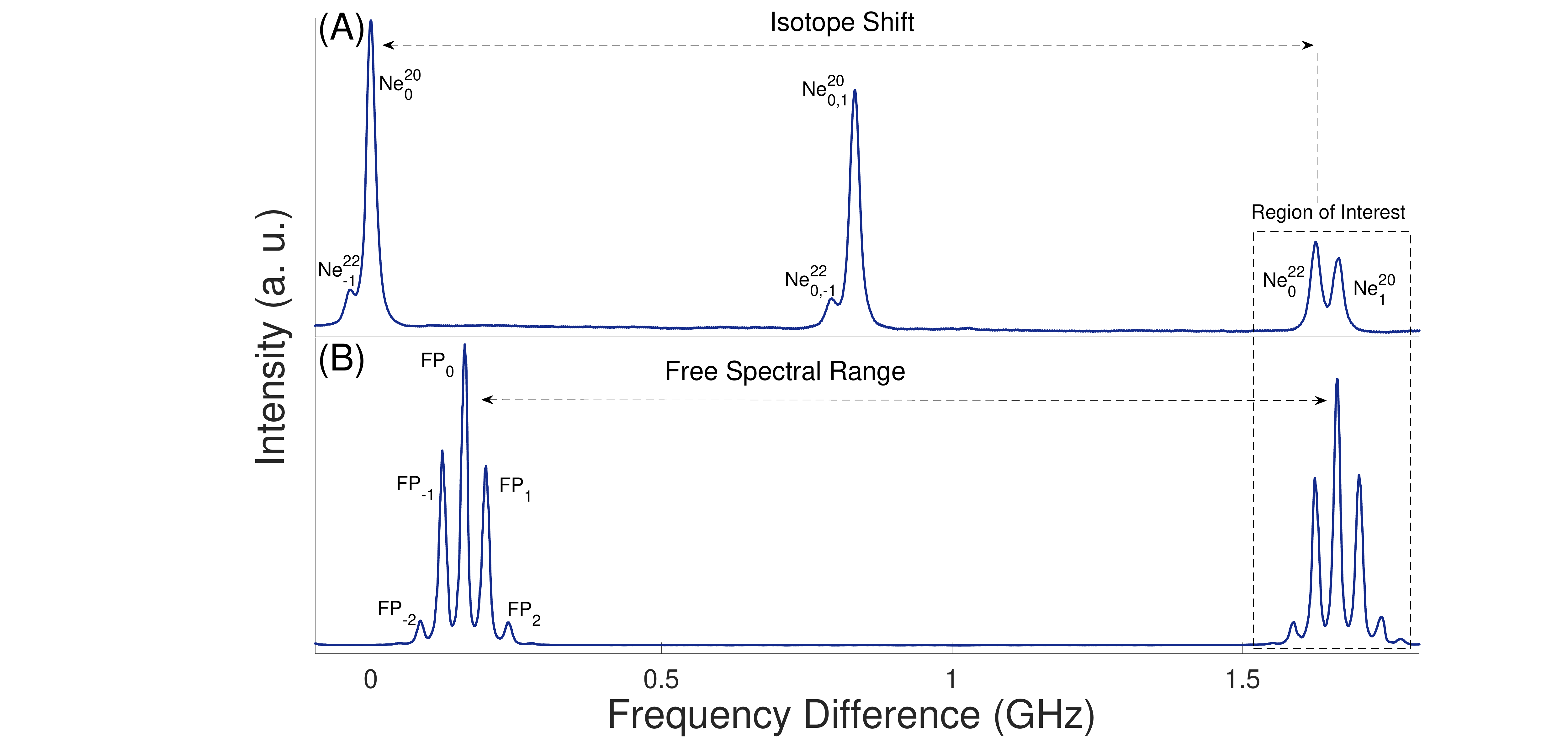}
	\caption{\label{Fig. Wide_lockin}
		Wide scan of laser frequency with phase modulations.
		(A) SA Lock-in signal, with $\omega_\mathrm{c}=$\SI{4}{kHz}, $\omega_\mathrm{M}=$\SI{1.7}{GHz}. Ne$^{X}_a$ denotes the $a$ sideband of isotope $X$. Crossovers between sidebands $a$,$a'$ are marked Ne$^{X}_{a,a'}$. 							
		(B) FP signal with $\omega_\mathrm{m}=$\SI{40}{MHz}. The FSR and sidebands are denoted.}
\end{figure}
with $\mathrm{J}_a=\mathrm{J}_a[m]$ the Bessel function of order $a$ with modulation index $m$, and the sideband Lorenzians $\mathrm{L}_i^a[\Delta]=\mathrm{L}_i[\Delta-a\omega_\mathrm{M}]$. Since they are independent of the laser frequency, we do not write the expressions for the peak amplitudes $\alpha_{i}^{k}$ explicitly. The second term in  (\ref{crossovers}) represents crossover peaks for each isotope obtained when the atoms are pumped by one sideband, and probed by another \cite{2008-EOM-Crossing}, $\mathrm{L}_{i}^{ab}[\Delta]=\mathrm{L}_i[\Delta-(a+b)\omega_\mathrm{M}/2]$. There are no crossover peaks between different isotopes.
Figure \ref{Fig. Wide_lockin} shows the measured atomic signal presented in  (\ref{crossovers}). We note that crossovers either fall between, or directly add, to the original peaks.

\section{Frequency calibration method}

In principle, it is possible to perform a wide scan similar to that presented in figure \ref{Fig. Wide_lockin}a and use the sideband peaks as  markers for calibration of frequency axis \cite{1997-EOM-freqcal}; however, a wide scan is more prone to frequency drifts and relies on either a completely linear scan or a complete determination of the nonlinearity \cite{2003-Li-EOM}. 
Instead, We scan the laser frequency only a small fraction of the actual separation, and calibrate the frequency axis using another modulated beam. When scanning the laser close to the second isotope resonance $\Delta\approx\omega_{\mathrm{IS}}$, and for a modulation frequency close to the isotope shift, $\omega_\mathrm{M}\approx\omega_{\mathrm{IS}}$ (region of interest in figure \ref{Fig. Wide_lockin}), only two peaks survive, which are separated by the difference between the modulation frequency and the isotope shift \begin{equation}\label{Signal_Fit}
\mathrm{V}[\Delta] \propto \mathrm{J}_1^2\alpha_1^1\mathrm{L}[\Delta-\omega_\mathrm{M}]+\mathrm{J}_0^2\alpha_2^0\mathrm{L}[\Delta-\omega_{\mathrm{IS}}].
\end{equation}
\begin{figure}[b]
	\centering
	\includegraphics[clip,trim={90 20 320 35 },width=\linewidth]{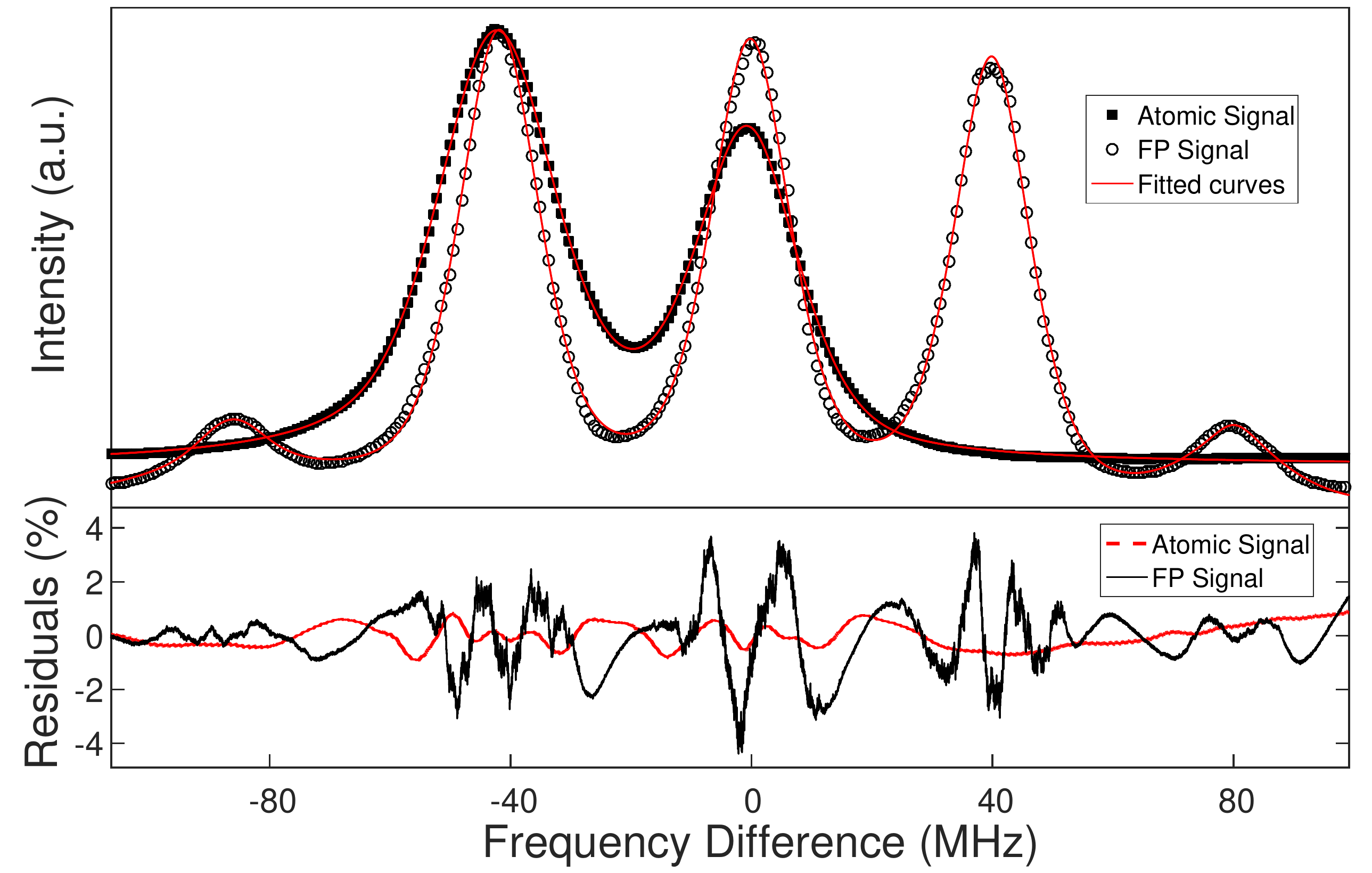}
	\caption{\label{Fig. Signal}
		Narrow scan of the region of interest (see Fig. \ref{Fig. Wide_lockin} for a wider scan), with typical EOM frequencies $\omega_m=40$ MHz and $\omega_M=1666$ MHz.
		Circle markers correspond to the atomic signal from the lock-in amplifier, fitted with (\ref{Signal_Fit}).  
		Full squares correspond to FP signal, fitted with (\ref{FP_signal}). Residuals are quoted as percentage of signal height. }
\end{figure}
To have the remaining peaks at a similar size, we choose the appropriate modulation index ($m\approx1)$. Generally, for a small peak with amplitude $\alpha_2$, and a larger one with $\alpha_1$, and since the modulation index can be arbitrarily small, one can always choose $m$ such that $(\mathrm{J}_1/\mathrm{J}_0)^2 \approx \alpha_2/\alpha_1$.
$(\mathrm{J}_1/\mathrm{J}_0)^2\approx (m/2)^2 \approx \alpha_2/\alpha_1$.

To accurately calibrate the frequency axis we split another beam, modulate its phase by $\omega_\mathrm{m}$, and insert it into a Fabri-P\'erot (FP) interferometer with finesse $F$ and FSR $\omega_{\mathrm{FSR}}$. 
The transmitted intensity can be written as \cite{2010-FP-FSR-EOM}
\begin{equation}\label{FP_signal}
\mathrm{I}[\Delta]=I_0\sum_{a=-\infty}^{\infty}{\mathrm{J}_a^2\mathrm{L}^a[\Delta,\omega_{\mathrm{FSR}}/F] },
\end{equation}
after filtering out terms oscillating at $a\omega_\mathrm{m}$ for $a\neq0$. Equation (\ref{FP_signal}) describes a series of Lorentzians, one for each sideband, separated by the modulation frequency. 
Figure \ref{Fig. Signal} shows the Lock-in signal in the region of interest, fitted with  (\ref{Signal_Fit}), along with the frequency calibration signal, fitted with  (\ref{FP_signal}).

\section{Implementation}
\begin{figure}[b]
	\includegraphics[clip,trim={60 87 20 50 },width=\linewidth]{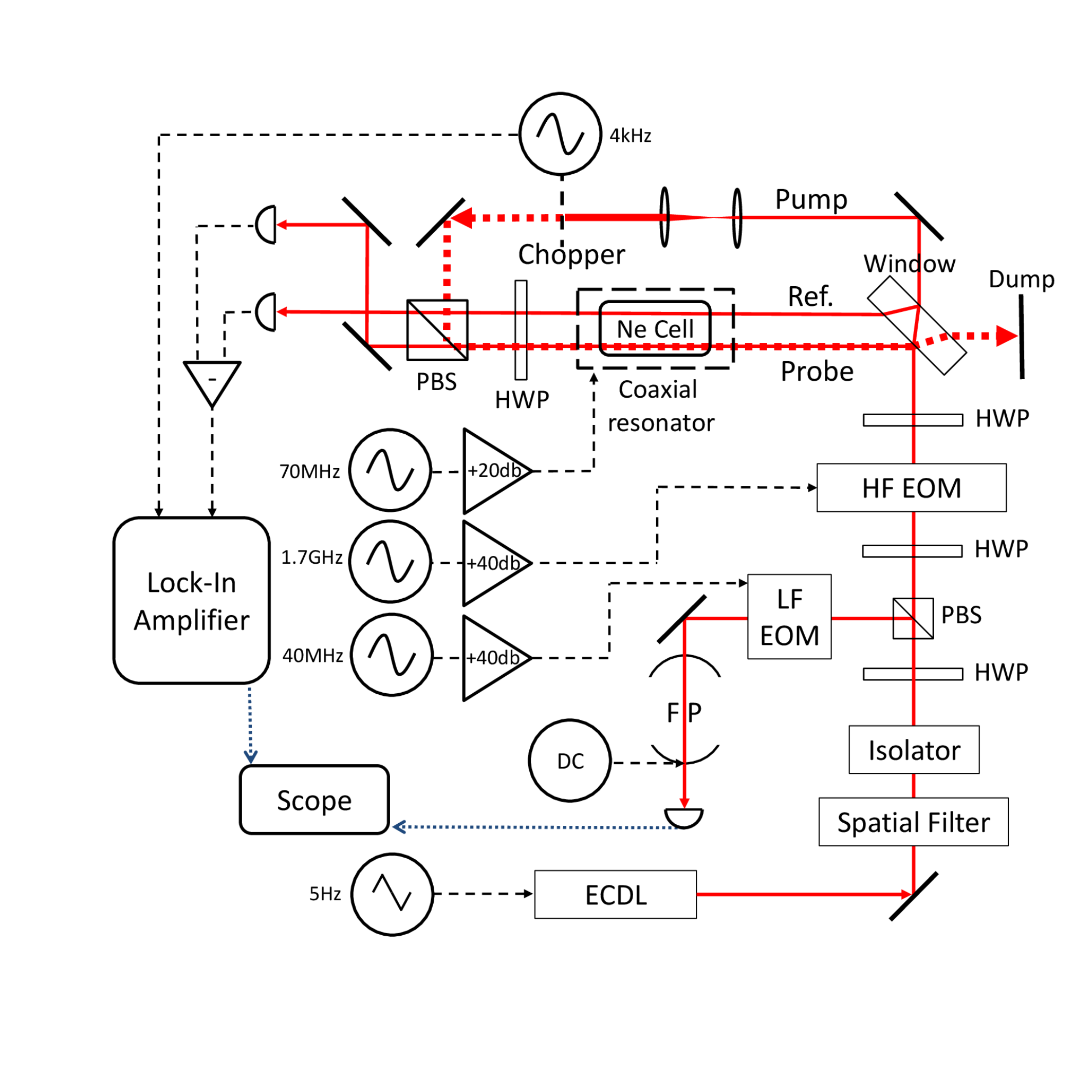}
	\caption{\label{Fig. Setup}
		Main elements of the experimental setup. EOM - Electro-optic modulator, HF - High frequency, LF - Low frequency, HWP - Half wave plate. FP - Fabri-P\'erot interferometer, ECDL - External cavity diode laser, PBS - Polartizing beam splitter. For pressure-dependent measurements (figure...), the sealed cell was replaced by a glass tube with gas inlet (see text).}
\end{figure}
We use a narrow-band ($<$MHz) single frequency laser beam, from a home-built external cavity diode laser \cite{2012-Steck}. Frequency scanning is performed by applying voltage to a piezoelectric element connected to the external cavity grating. The beam is split in two. One part goes through a broadband, low frequency (DC-\SI{100}{MHz}) commercial electro-optic-modulator (EOM, New Focus 4002) and into a Fabri-P\'erot cavity (Thorlabs SA-200, $\omega_{\mathrm{FSR}}=$\SI{1.5}{GHz}, $F=200$). The other part goes through a home-built, narrowband, high frequency EOM \cite{1987-EOM}, and enters a collinear, linearly polarized
, pump-probe type setup with high-purity, natural abundance neon gas (90\% $^{20}$Ne, 9\% $^{22}$Ne and 0.3\% $^{21}$Ne), contained in a AR-coated, glass cell, which resides in a high-Q coaxial resonator \cite{1959_resonator}. An RF-driven discharge at the resonance frequency (\SI{70}{MHz}) excites the atoms and populates higher lying states. After ignition, a few milliwatts of RF-power are sufficient to maintain stable plasma.
The pump beam is amplitude modulated by a chopper at $\omega_\mathrm{c}=$ \SI{4}{kHz}. A reference beam goes through the cell as well, and provides another stage of subtraction to remove amplitude noise resulting from the laser (in part due to pointing instability and birefringent effects in the EOM) and cell discharge.  The signal is fed into a lock-in amplifier (SRS SR830) where it is mixed with the chopper reference,  filtered and amplified.
Figure \ref{Fig. Setup} shows the main elements of the experimental system.	

A slow (few Hz) and narrow (\SI{200}{MHz}) scan of the laser frequency results in traces of the lock-in and FP signals simultaneously (figures \ref{Fig. Wide_lockin} and \ref{Fig. Signal}). We tune the relative FP frequency position by applying DC voltage to a piezoelectric element moving one of the cavity mirrors. From  (\ref{Signal_Fit}), the distance between the zero-order $^{22}$Ne peak and the first-order $^{20}$Ne is exactly $\omega_\mathrm{d}=\omega_\mathrm{M}-\omega_{\mathrm{IS}}$. We tune the low-frequency EOM to $\omega_\mathrm{m}\approx\omega_\mathrm{d}$ by placing two of the FP sideband peaks directly on top of the lock-in atomic signal peaks (see figure \ref{Fig. Signal}). This limits the effects of scan nonlinearity in calibration of the frequency axis to less than a few kHz per trace. To each trace we fit the atomic signal with two Lorentzians of  (\ref{Signal_Fit}), and the FP signal with five Lorentzians corresponding to the $0,\pm1,\pm2$ sideband orders observed (\ref{FP_signal}). To account for non-homogeneous broadening, and so model the tails of the peaks accurately, each Lorentzian in the fits is replaced with a pseudo-Voigt profile \cite{2012_Voigt}.
The fitting procedure gives the distances between the FP peaks $\tau_{\mathrm{FP}}$ and the Atomic peaks $\tau_{\mathrm{LI}}$ in units of time, and so the isotope shift is calculated as:
\begin{equation}\label{shift}
\omega_{\mathrm{IS}}=\omega_\mathrm{M}-\omega_\mathrm{d}=\omega_\mathrm{M}-\omega_\mathrm{m}\frac{\tau_{\mathrm{LI}}}{\tau_{\mathrm{FP}} }.
\end{equation}

This procedure of obtaining the IS is robust against frequency drifts in the laser (few MHz per minute), since both the atomic and FP signals drift together. The FP FSR is not used, and so slow (MHz per several minutes) thermal drifts in the cavity length only serve to move the FP signal relative to the atomic signal. 

\section{Results and discussion}
\begin{figure*}[htbp]
	\centering
	\includegraphics[clip,trim={0 2 8 12 },width=\textwidth]{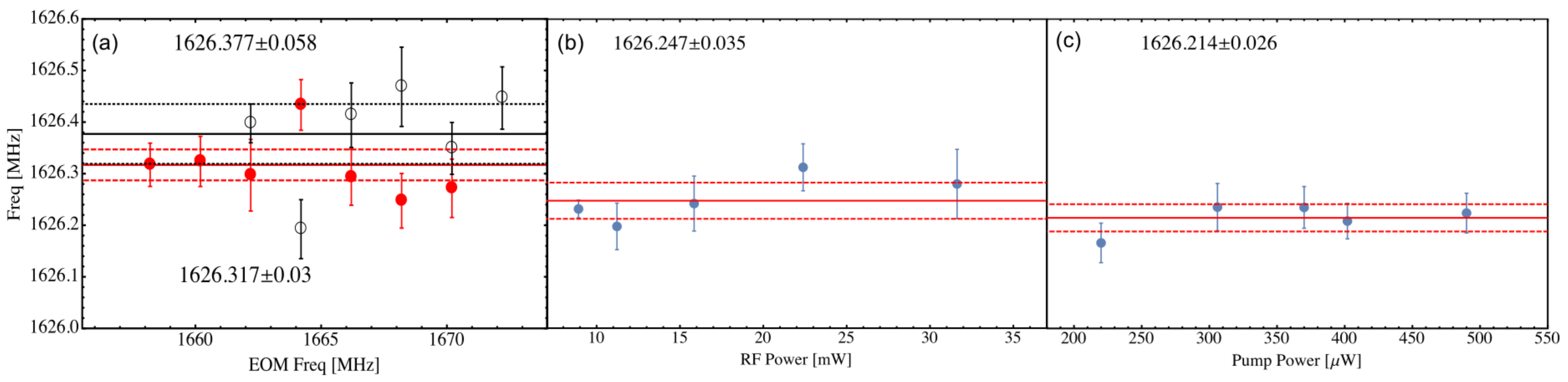}
	\caption{\label{Fig. results} Calculated isotope shift frequency when varying experimental parameters, and under a constant pressure of \SI{200}{mTorr}. Bold horizontal lines are weighted average and dashed lines represent confidence bounds of $68\%$.
		(a) Changing both EOM frequencies (see text). Horizontal axis labeled by the narrowband EOM frequency. Full circles are at $10$ mW RF power and empty circles are at $30$ mW.
		(b) Changing RF power.
		(c) Changing pump laser power.}
\end{figure*}
For each experimental run, about $20$ traces are taken with identical parameters (laser power, pressure, etc.). The results are calculated 
using  (\ref{shift}), and averaged using a Bayesian analysis approach with the \emph{Just Another Gibbs Sampler} (JAGS) program \cite{2010-andreon_scaling, 2016-sereno_bayesian}, which takes into account possible correlations between measurement errors and their intrinsic scatter.
\begin{figure}[b]
	\centering
	\includegraphics[clip,trim={150 20 220 30 },width=\linewidth]{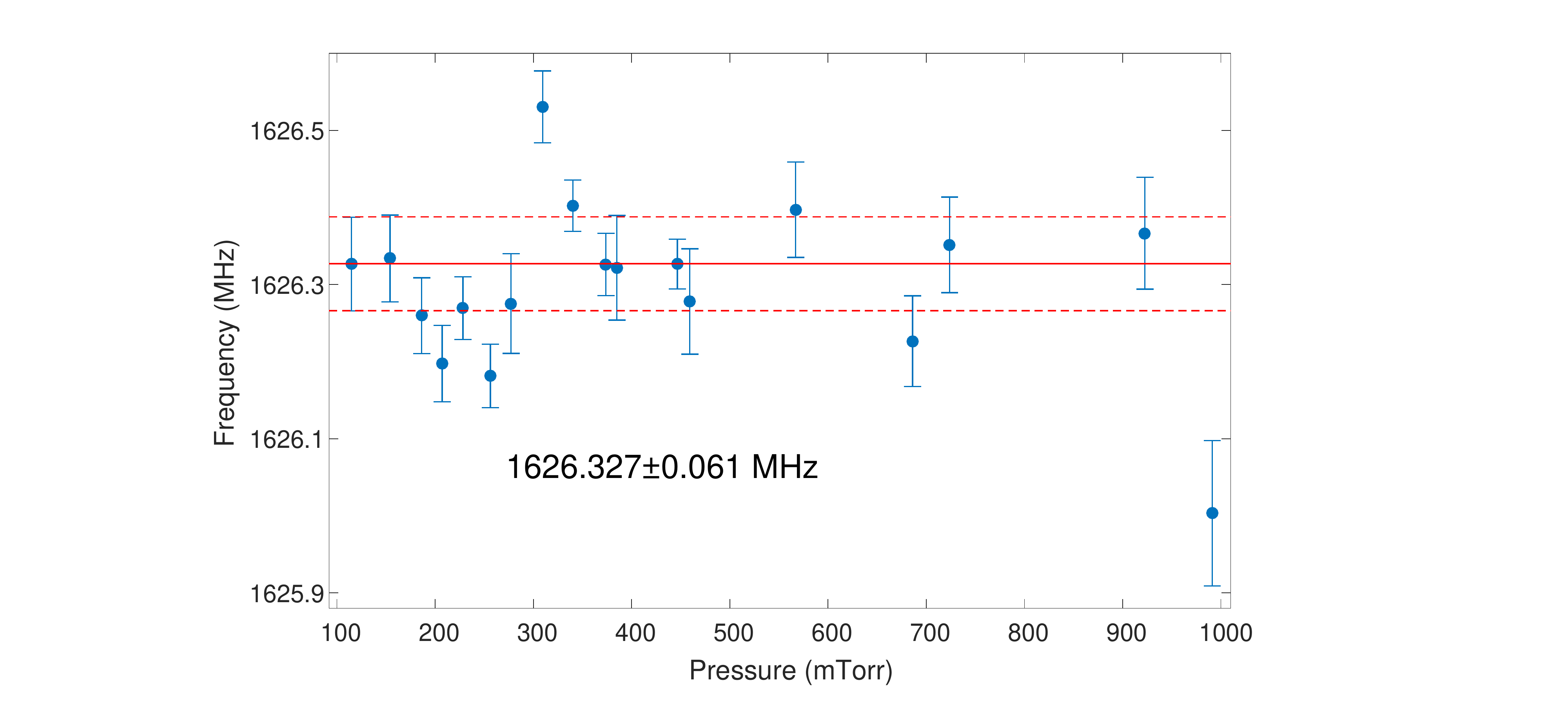}
	\caption{\label{Fig. results2} Calculated isotope shift frequency for varying pressure. Wighted confidence bounds to $68\%$ are indicated.}
\end{figure}

To account for unknown systematic effects we investigate the calculated IS for different experimental parameters. By varying the laser power (figure \ref{Fig. results}c), we change the width of the peaks through saturation broadening \cite{2013-Pressure}, and their height. By varying the RF-discharge power (figure \ref{Fig. results}b), we change the excited-state population and peak height, as well as shifts which may result from non-thermal distribution of the gas sample. Hysteretic effects were observed at high RF power, where coupling of the radio-waves to the plasma changed from capacitive to inductive \cite{2015-RF_source}, and so we limited our investigation to low powers. The most stringent test for our measurement scheme is to vary both EOM frequencies together (figure \ref{Fig. results}a). This changes both the distance and magnitude of all peaks involved.
The above measurements were done with a sealed cell at a pressure of \SI{200}{mTorr}. We then replaced it with a glass tube that has a gas inlet. The tube was first pumped to under one mTorr and then filled with high purity, natural abundance neon gas at various pressures. The pressure reading was stable to better than $1\%$ during an experimental run. The results of this set are presented in figure \ref{Fig. results2}. Even though similar lines for $^{20}$Ne are expected to shift by about \SI{2}{MHz/Torr} \cite{1996-Leo_press_neon}, no pressure shift in the IS was observed within our measurement uncertainties, which indicated that the shift is similar between the isotopes to a few ten kHz per Torr.

The results of the sets presented in figure \ref{Fig. results} and \ref{Fig. results2} are combined using the JAGS framework to obtain a wighted result of $1626.289\pm$\SI{53}{MHz}, where the quoted uncertainty range is one standard deviation.
\begin{figure}[b]
	\includegraphics[clip,trim={26 22 15 1 },width=\linewidth]{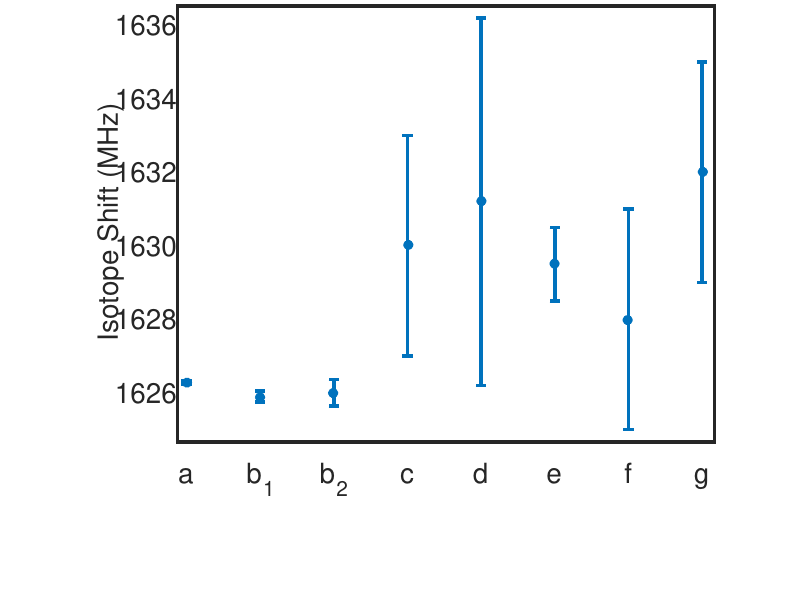}
	\caption{\label{Fig. compare}
		Isotope shift and $68\%$ confidence intervals obtained by different groups: 
		(a) This work,
		(b) Feldker \textit{et. al.} \cite{2011-Birkl_IS}, 1 - Absorption, 2 - Fluoresence,
		(c) Julien \textit{et. al.} \cite{1980-Julien},
		(d) Guth\"ohrlein \textit{et. al.} \cite{1994-Guth},
		(e) Basar \textit{et. al.} \cite{1997-Bassar-NeI_lines},
		(f) Odintsov \textit{et. al.} \cite{1965-Odinstov}.
		(g) Konz \textit{et. al.} \cite{1992-Konz-Craft_spectroscopy}.
	}
\end{figure}
\begin{table*}
	\centering%
	\begin{tabular}{|c|c|c|}
		\hline 
		Reference & Reported Value (MHz) & Method \\ 
		\hline 
		This Work & $1626.287\pm0.053$ & Dual-sideband saturated absorption \\ 
		\hline 
		\cite{2011-Birkl_IS} & $1625.9\pm0.15$ & Trap absorption \\ 
		\hline 
		\cite{2011-Birkl_IS} &  $1626.0\pm0.22$ & Trap fluorescence \\ 
		\hline 
		\cite{1980-Julien} & $1630\pm3$ & Velocity selective optical pumping \\ 
		\hline 
		\cite{1994-Guth} & $1631.2\pm5.0$ & Optogalvanic spectroscopy \\ 
		\hline 
		\cite{1997-Bassar-NeI_lines} & $1629.5\pm1.0$ &  Intermodulated optogalvanic spectroscopy \\ 
		\hline 
		\cite{1965-Odinstov} & $1628\pm3$ & Doppler-free two-photon spectroscopy \\ 
		\hline 
		\cite{1992-Konz-Craft_spectroscopy} & $1632\pm3$ & Supersonic Beam \\ 
		\hline 
	\end{tabular} 
	\caption{Isotope shift and standard error obtained by various experimental techniques.}\label{table:somename}
\end{table*}

We now discuss the contributions of some known systematic corrections, which are not affected by the parameters scanned, to the obtained experimental value. We note here that in our measurement scheme, the $^{22}$Ne peak appears at a lower laser frequency than the $^{20}$Ne peak (See figure \ref{Fig. Wide_lockin}).
Due to their similar electronic configuration and identical quantum numbers, most of the systematic shifts between the lines of $^{20}$Ne and $^{22}$Ne vanish to high orders when measuring the isotope shift. Among those are Zeeman shifts.
The $^3\mathrm{P}_2$ and $^3\mathrm{D}_3$ levels in neon are 24 and \SI{5} {THz} away from their closest neighbours respectively. Since quantum interference shift is inversely proportional to the difference between the levels \cite{2012-Quantum-Interf-Theory}, this effect is vanishingly small in our case.
Naturally, the main difference between the two isotopes is their mass $M$. It affects the atomic recoil to create the so-called recoil shift of $\omega_\mathrm{r}=h/(2M\lambda^2)$, a \SI{-2.2}{kHz} shift to the IS. The thermal distribution cancels first order Doppler shifts but adds a second order shift of $-4T/(\lambda c\pi M)$ \cite{2013-Pressure}, a negligible \SI{75}{Hz} correction.
The corrected result for the isotope shift is thus $1626.287\pm$\SI{53}{MHz}.

\section{Conclusion}

We presented a simple measurement scheme for accurate determination of intervals between far (up to few GHz) lying resonances in a spectroscopy signal. This method was used to measure the isotope shift between the $^{20}$Ne and $^{22}$Ne cooling transition with high precision.
Figure \ref{Fig. compare} and table \ref{table:somename} shows a comparison between the results presented here, and those of other groups using various experimental techniques. We note that earlier attempts \cite{1980-Julien,1994-Guth,1997-Bassar-NeI_lines,1965-Odinstov,1992-Konz-Craft_spectroscopy}, obtain a \SI{4}{MHz} larger shift than more recent and accurate ones presented in this work and in \cite{2011-Birkl_IS}. It would thus be beneficial to conduct a high accuracy, ab initio calculation of this quantity, which as far as we know, does not exist in the literature \cite{2011-Birkl_IS}.

To check our results with a different experimental system, we intend to conduct this measurement in our trap setup \cite{2015-Our-MOT}. By measuring $^{21}$Ne as well, it is also possible to improve determination of the $^{20-22}$Ne charge radii difference \cite{2008-Blaum_Ne_mass_radius}.

This work was supported by the Israeli Science Foundation under ISF Grant No. (139/15);
B.O. is supported by the Hoffman leadership and responsibility program, and the Eshkol Fellowship of the Ministry of Science and Technology.

\bibliographystyle{iopart-num}
\bibliography{All}
\end{document}